\documentclass[aps,prl,reprint,groupedaddress]{revtex4-1}
\usepackage{amsmath, graphicx, verbatim}

\begin{document}

\title{Sliding elastic lattice: an explanation of the motion of superconducting vortices}

\author{Young-noh Yoon$^{1*}$ and Jonghee Lee$^2$}
\email{mystyle@umd.edu; jonghee@umd.edu}
\affiliation{$^1$Department of Physics, University of Maryland, College Park, Maryland 20742, USA\\$^2$Department of Materials Science and Engineering, University of Maryland, College Park, Maryland 20742, USA}
\date{\today}

\begin{abstract}
We introduce a system where an elastic lattice of particles is moved slowly at a constant velocity under the influence of a local external potential, construct a rigid-body model through simplification processes, and show that the two systems produce similar results. Then, we apply our model to a superconducting vortex system and produce path patterns similar to the ones reported in [Lee et~al., Phys.\ Rev.\ B \textbf{84}, 060515 (2011)] suggesting that the reasoning of the simplification processes in this paper can be a possible explanation of the experimentally observed phenomenon.
\end{abstract}


\maketitle

The systems having many interacting particles, such as superconducting vortices \cite{Abrikosov-57}, charge density waves \cite{Gruner-88}, colloids \cite{Korda-02}, and Wigner crystals \cite{Wigner-34}, have been of great interest in condensed matter physics because they show rich dynamic and static properties and the various phases that originate from the nonequilibrium dynamics and statistics. For example, the motion of superconducting vortices under the influence of pinning potentials has been extensively studied in experiments \cite{Harada-et-al-92, Moser-et-al-95, Pardo-et-al-98, Troyanovski-et-al-99, Olsen-et-al-04, Guillamon-et-al-09, Kalisky-et-al-09}, theories \cite{Giamarchi-and-Bhattacharya-01, Blatter-et-al-94}, and simulations \cite{Olson-et-al-98, Reichhardt-et-al-00}, because the vortex system can be a model for the study of the dynamics of many interacting particle systems. Moreover, understanding the motion of vortices is important for the development of the superconductivity-based devices \cite{Lee-et-al-99}. One of the main topics of the vortex dynamics study is how to effectively suppress the motion of vortices because the heat dissipation caused by the motion of vortices can destroy the superconducting state. Therefore, extensive studies \cite{Giamarchi-and-Bhattacharya-01, Blatter-et-al-94, Harada-et-al-92, Moser-et-al-95, Pardo-et-al-98, Troyanovski-et-al-99, Olsen-et-al-04, Guillamon-et-al-09, Kalisky-et-al-09, Olson-et-al-98, Reichhardt-et-al-00} have focused on the pinning effects on vortex motions when the vortex-pinning interactions are large enough to disturb the regularity of the Abrikosov vortex lattices \cite{Abrikosov-57}. However, vortex dynamics under weak pinning potentials has been rarely studied to date. Here, we ask ourselves, how do the moving vortices behave when the vortex-vortex interactions dominate the vortex-pinning interactions?

Recently, the motion of the superconducting vortices in a single crystal NbSe$_2$ sample was observed over 7 days using a scanning tunneling microscopy when the vortices were driven at a few pm/s by a slowly decaying magnetic field at the rate of a few nT/s \cite{Lee-et-al-11}. The paper emphasizes two peculiar features in the observed vortex dynamics: the nonuniform motion of the vortices and the existence of favored locations by the moving vortices. A 2-dimensional molecular dynamics simulation is a good candidate for the understanding of the observed features. However, it requires a long computation time with a large number of particles. Indeed, an extensive 2-dimensional molecular dynamics simulation with about 8000 particles has been done \cite{Dreyer-et-al-09}, but the simulation was not able to provide a physical insight into the observed peculiar features due to the limitation of the computing power. In this paper, we propose a model that can provide a good physical insight into the origin of the observed features without use of extensive computing power. 
  
The first goal of this paper is to study the properties of a 2-dimensional elastic lattice and construct a computationally efficient approximate model that captures the basic mechanism of the behavior of the slowly moving elastic lattice under the influence of a weak local attractive potential fixed in space. The second goal of the paper is to show that this model can be applied to a superconducting vortex system under certain conditions and produce similar result, thus providing a possible explanation of the experimentally observed features of superconducting vortices reported in \cite{Lee-et-al-11}.

A superconducting sample develops current vortices when an external magnetic field is applied to the sample perpendicular to the surface of the sample. The vortices exert repulsive forces among them due to electromagnetic forces. Vortices can be treated as classical particles with repulsive forces among them packed inside the boundary of the sample. They form a triangular lattice structure when the external magnetic field is strong enough. As the strength of the magnetic field decreases, the vortices near the boundary of the sample disappear and the number of vortices in the sample decreases making the distances among them increase. This causes vortex motion roughly in radial directions away from the center of the sample. It was reported that the vortices move non-uniformly at the speed on the order of pm/s with the triangular lattice structure maintained in the scanning window and there are favored locations where vortices visit and stay longer than at other locations \cite{Lee-et-al-11}.

To make a model that produces the observed behaviors, we assume a quasistatic equilibrium because the vortices move extremely slowly. We also assume there are impurities in the sample introducing external attractive potentials and that they are weak because of the pristine nature of the sample. We are interested in the motion of the vortices near the impurity close to the scanning area. To estimate the computational cost of the direct simulation, consider particles between 2 fixed boundary points at the origin and $L$ in a 1-dimensional space and the motion of the particles as the particles next to the right boundary are removed one by one. Removing 1 particle causes a particle at $x$ move about $x/L$ of the lattice constant. The observation location needs to be less than $L/10$ so that the paths of the particles at the observation location have at least 10 steps as the particles move for about 1 lattice constant. We need to remove at least 10 particles so that the particles at the observation location move for at least 1 lattice constant. We need to have at least 100 initial particles so that the lattice constant does not change much as reported in \cite{Lee-et-al-11} as 10 particles are being removed. In a 2-dimensional case, we need to have at least $100\times100$ particles making the direct simulation computationally too expensive to perform.

We note that the direct cause of the motion of the vortices is the moving neighboring vortices. If we consider a collection of vortices of a certain area, then the motion of those vortices is caused by the repulsive forces from the surrounding moving vortices. Replacing the surrounding vortices with a repulsive boundary moving at a constant velocity to approximate the driving mechanism and introducing a weak external attractive potential fixed in space for the effect of the weak pinning caused by an impurity leads to a moving boundary model. However, this model did not produce the desired features possibly because we could not simulate enough particles limited by computing power, assuming the reported features in \cite{Lee-et-al-11} are the results of a large number of vortices. This assumption is explained later in the simplification process from the elastic-lattice model to the rigid-body model.

Assuming the pinning is weak enough not to break the lattice structure of the vortices, we replace the particles with repulsive forces in the previous model with a lattice of particles and springs. This motivates the sliding elastic lattice model. Consider a 2-dimensional triangular lattice structure of particles with each connected to 6 neighbors by springs of natural length 1 inside a certain boundary under the influence of a weak local external potential at a fixed location. We drag the boundary at a constant velocity slowly enough to ignore the masses of the particles and use quasistatic equilibrium analysis. We assume that the external potential is weak enough not to distort the lattice structure too much and that the external potential is short-ranged compared to the unit length of the lattice. We are interested in the paths of the particles near the external potential.

Even if the elastic lattice model is computationally more efficient than the moving boundary model, the computational cost of the elastic lattice model is still high. To make a computationally more efficient model, we draw a virtual boundary containing several particles of our interest near the external potential resulting in the system divided into the area of our interest and the bigger environment containing the rest of the particles. There are three orders of strength involved in this picture. Firstly, the effective spring constant of the environment is much less than the spring constant of the individual springs because the environment contains many springs connected in part in series and the combined spring constant decreases when springs are connected in series. Secondly, the strength of the external potential is weak enough not to distort the lattice structure much, if we assume the external potential is weak compared to the strengths of the individual springs. Thirdly, the external potential is strong enough to disturb the locations of the particles of our interest because the effective spring of the environment is soft. Thus, we expect that physics does not change much when we replace the springs and the particles inside the virtual boundary with a rigid body of those particles and replace the environment with a soft representative spring connected to the rigid body.

We need to characterize the properties of the lattice before we simplify the system. To do that, we consider an elastic lattice with one particle located at the origin and the triangular structure oriented so that one of the 6 directions of the edges is parallel to the $x$-axis, ignoring the existence of the external potential at the moment. We cut out the lattice outside the circle of radius $r$ centered at the origin when the lattice is at equilibrium with lattice constant 1. We call the particles that have all 6 neighboring particles inside particles, and the particles that have less than 6 neighboring particles boundary particles. We fix the relative positions of the boundary particles to each other, letting the inside particles free to move.

\begin{figure}
\includegraphics{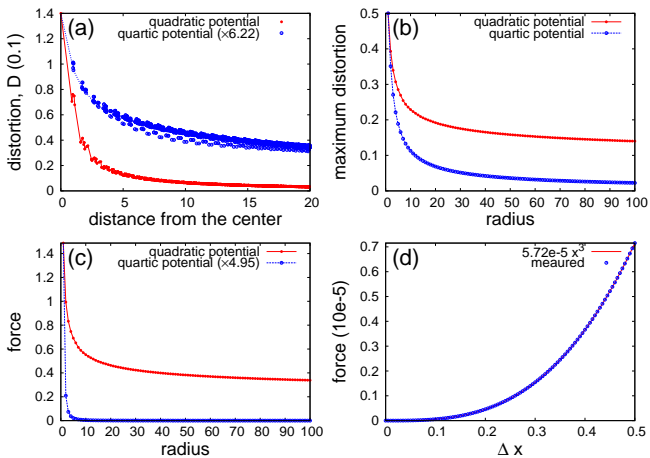}
\caption{(color online) (a) Distribution of distortions. The radius of the boundary is 100. The values with quartic potential springs were multiplied by 6.22. (b) Maximum distortion vs.\ lattice size (radius). The vertical axis is the maximum distortion occurring at the center particle moved by the external force. (c) force vs.\ lattice size (radius). The vertical axis is the force necessary to move the center particle by 0.5 distance. The forces with quartic potential springs were multiplied by 4.95. (d) Force vs.\ $\Delta x$. The quartic potential springs are used. The radius of the lattice is 100. The hollow blue dots are the measured forces necessary to move the center particle by $\Delta x$. The solid curve is a fit by cubic function $5.72 \times 10^{-5} x^3$.}\label{characterization}
\end{figure}

A 2-dimensional lattice is somewhat different from a 1-dimensional lattice, because the 2-dimensional lattice becomes unevenly distorted when the center particle is moved by an external force while the 1-dimensional lattice becomes evenly compressed or stretched resulting in the effective spring constant of the system being inversely proportional to the size of the system. Thus, we need to characterize the distortion of the lattice. To characterize the distortion, we moved the center particle of the lattice to the right by 0.5 distance keeping the boundary particles at the same positions. We found that the resulting displacements of the particles are quite local near the center, and the distortion of the hexagon of the 6 neighboring particles surrounding the center one is not negligible. Thus, replacing the lattice near the center one with a rigid body is not reasonable in this situation. But if the functional form of the potential energy of each individual spring has a shape that introduces more penalty away from the equilibrium than a quadratic function (obeying Hooke's law) does, then we can expect that the displacements become more spread and the distortion near the center decreases. To quantify the distortion, we define distortion $D_n$ of the hexagon surrounding the $n$-th particle as
\begin{equation}
D_n^2 = \frac{\sum_{i=1}^{6} \left[ \, (x_{n,i}-x'_{n,i})^2 + (y_{n,i}-y'_{n,i})^2 \, \right]}{6},
\end{equation}
where $x_{n,i}$ and $y_{n,i}$ are the $x$ and $y$ positions of the $i$th neighboring particle of the distorted hexagon and $x'_{n,i}$ and $y'_{n,i}$ are the $x$ and $y$ positions of the $i$th neighboring particle of an undistorted hexagon. Figure~\ref{characterization}(a) shows the distortions as a function of distance from the center with springs of $V=\frac{1}{2}(\Delta x)^2$ potential energy (quadratic) and $V=\frac{1}{4}(\Delta x)^4$ potential energy (quartic) when the radius of the boundary is 100. The figure shows that the distortions are more spread throughout the lattice and the maximum distortion occurring at the center is smaller with quartic potential springs than with quadratic potential springs. Figure~\ref{characterization}(b) shows the maximum distortion, which occurs at the center particle that was moved by the external force, as a function of the lattice size (the radius of the boundary) with quadratic potential springs and with quartic potential springs. The figure shows that the maximum distortion decreases faster with quartic potential springs than with quadratic potential springs as the lattice size increases implying the rigid body model is a better approximation with quartic potential springs than with quadratic potential springs given the same lattice size.

Next, we characterize the effective spring of the lattice. Figure~\ref{characterization}(c) shows the amount of external force necessary to move the center particle by 0.5 distance as a function of the lattice size with quadratic potential springs and quartic potential springs. The figure shows that the force necessary to move the center particle decreases much faster with quartic potential springs than with quadratic potential springs as the lattice size increases, and equivalently that the effective spring the lattice becomes softer much faster with quartic potential springs than with quadratic potential springs as the lattice size increases.

So far, we compared the properties of the lattice of quadratic springs and the lattice of quartic springs. We also found that the results extend further when the higher exponents than 2 or 4 are used in the potential function of the springs. For the use in the rigid body model, we measured the spring constant $k$ of the lattice of quartic springs when the radius of the lattice is 100 by measuring the forces necessary to move the center particle by $\Delta x$ ranging from 0 to 0.5. Figure~\ref{characterization}(d) shows the force as a function of $\Delta x$. We find the spring constant $k$ of the lattice to be $5.72 \times 10^{-5}$ from the figure.

\begin{figure}
\includegraphics{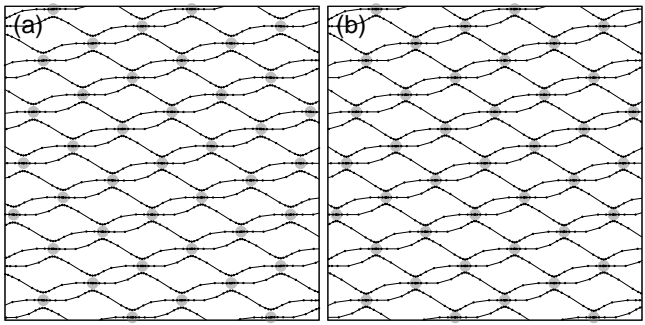}
\caption{Paths of the particles near the external potential of (a) the elastic lattice model (b) the rigid body model when the system moves to the right in small intervals with the external potential located at the origin. The small black dots are the positions of the particles at each distance step, and the big gray dots are the positions of the particles with the center particle at the origin and no external potential.}\label{path}
\end{figure}

Now, we compare the elastic lattice model and the rigid body model. The elastic lattice model is the same as described before with a lattice of the radius of 100, springs of $V=\frac{1}{4}(\Delta x)^4$, and an external potential of 2-dimensional inverted Gaussian with standard deviation $w/\sqrt{2}$ of $0.2/\sqrt{2}$ and height $h$ of $1.144 \times 10^{-7}$ fixed at the origin. We rotated the lattice by arctan$(\sqrt{3}/5)$ radian counterclockwise to produce a nontrivial path pattern, moved the boundary particles to the right in small intervals keeping the shape of the boundary the same but allowing the inside particles move freely, and then numerically calculated the equilibrium positions of the particles at each step by finding the positions of the inside particles that minimize the total potential energy defined as
\begin{equation}
V(\mathbf{x}, \mathbf{y}) = \frac{1}{4} \sum_{n=1}^{N_s} (l_{n}-1)^{4} - \sum_{n=1}^{N} h \, e^{-(x_{n}^{2}+y_{n}^{2})/w^2},
\end{equation}
where $\mathbf{x}$ and $\mathbf{y}$ are vectors with each component representing $x$ and $y$ positions of each particle, $N_s$ is the number of springs of the system, and $l_n$ is the length of the $n$th spring that is determined by the positions of the two particles on the ends, and $N$ is the number of the inside particles. Figure~\ref{path}(a) shows the path pattern near the external potential. The small black dots are the positions of the particles at each traveled distance by the boundary particles. The solid curves connecting the small black dots are the paths of each particle. The big gray dots are the positions of the particles with the center particle at the origin when there is no external potential. The figure shows that the particles tend to stay longer near the gray areas. We did a similar simulation with a rigid body of a rotated hexagonal shape of radius 20 and an effective spring of spring constant $k$ of $5.72 \times 10^{-5}$ as determined in Fig.~\ref{characterization}(d) attached to the center particle. We moved the equilibrium position of the effective spring to the right in small intervals, and calculated the position of the center particle that minimizes the total potential energy defined as
\begin{equation}
V(x,y) = \frac{k}{4} \left[(x-X)^2+y^2\right]^2 - \sum_{n=1}^{N} h \, e^{-\frac{(x+x_{n})^{2}+(y+y_{n})^{2}}{w^2}},
\end{equation}
where $x$ and $y$ are the horizontal and vertical positions of the center particle of the rigid body, $X$ is the horizontal equilibrium position of the center particle due to the effective spring, $N$ is the number of particles of the rigid body, and $x_n$ and $y_n$ are the fixed horizontal and vertical positions of the $n$th particle in the rigid hexagon with respect to the center particle. The second term of the potential can be viewed as the potential of the center particle due to multiple Gaussian potentials, so the system can be viewed as one particle attached to a spring and moving through multiple external potentials located at the gray areas in the figure. Figure~\ref{path}(b) shows the result. Figs.~\ref{path}(a) and \ref{path}(b) show similar path patterns implying that the rigid body model is a good approximation of the elastic lattice model.

\begin{figure}
\includegraphics[scale=0.95]{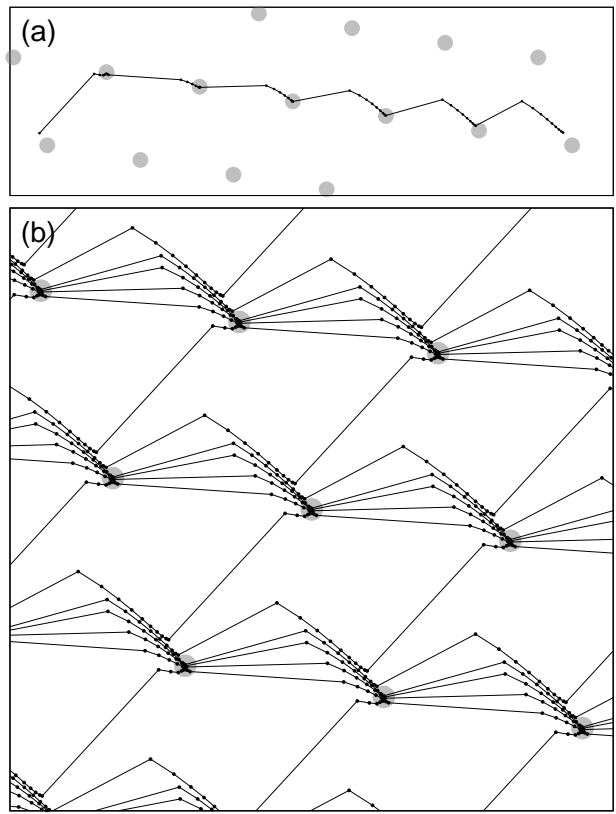}
\caption{The path patterns of the rigid body model. (a) The path of the center particle. (b) The overlaid paths of several particles around the center particle. The lattice was rotated by -arctan($\sqrt{3}/11$) radian. An asymmetric Gaussian external potential was used. The equilibrium position of the effective spring was moved to the left in small intervals.}\label{vortex}
\end{figure}

To produce the path pattern of the superconducting vortices reported in \cite{Lee-et-al-11}, we used a rigid hexagon of radius 20 rotated by -arctan($\sqrt{3}/11$) radian, a spring of spring constant 1 attached to the center particle, and an external potential defined as
\begin{align}
G(x, y) &=
\begin{cases}
	-A \, e^{-(u^2/\sigma_{u+}^2+v^2/\sigma_v^2)}, \textrm{ for } u > 0\\
	-A \, e^{-(u^2/\sigma_{u-}^2+v^2/\sigma_v^2)}, \textrm{ otherwise}
\end{cases}\\
u &= x \, \text{cos} \, \theta - y \, \text{sin} \, \theta \nonumber\\
v &= x \, \text{sin} \, \theta + y \, \text{cos} \, \theta, \nonumber
\end{align}
where $A = 0.03$, $\sigma_{u+}=0.6$, $\sigma_{u-}=\sigma_v=0.2$, and $\theta=-132^{\circ}$. We moved the equilibrium position of the spring to the left. Figure \ref{vortex} shows the path of the center particle and the overlaid paths of several particles around the center. This figure and Fig.~2 of \cite{Lee-et-al-11} show similar path patterns even if the paths from our model make sharper changes of directions than the paths from the experimental observation do. From this result, we propose that the pinning effects of the impurities in the sample used in the experiment of \cite{Lee-et-al-11} were weak enough not to destroy the lattice structure of the vortices around them, but the large number of vortices in the sample made the effective spring of the vortex system soft enough to produce the observed vortex dynamics.

In summary, we went through several simplification processes of the superconducting vortex system introducing the sliding elastic lattice model with a weak external potential and a rigid body model that approximates the behavior of the particles near the external potential, and showed similar path patterns from the elastic lattice model and the rigid body model. Then, we applied the rigid body model to the superconducting vortex system under a slowly decaying magnetic field, and showed that the path pattern from our model is similar to the experimentally observed one suggesting that the reasoning leading to our model can be a possible explanation of the observed vortex dynamics reported in \cite{Lee-et-al-11}.

\end{document}